\documentclass[aps,prl,preprint]{revtex4}
\usepackage{amsmath,amssymb}
\usepackage{graphicx,xcolor}
\begin{document}
\title{Optimal Clock Speed of Single-Qubit Operations on Open Quantum Systems}
\author{Nilanjana Chanda}
\email{nc16ip020@iiserkol.ac.in}
\author{Rangeet Bhattacharyya}
\email{rangeet@iiserkol.ac.in}
\affiliation{Department of Physical Sciences, Indian Institute of Science Education and Research Kolkata, 
Mohanpur -- 741246, WB, India}

\begin{abstract}

Efficient implementation of quantum algorithms requires single- or multi-qubit
gates with high fidelity.  Here, we report that the fidelity of single-qubit
gate operations on open quantum systems has a maximum value corresponding to an
optimum value of the drive-amplitude in the presence of drive-induced
decoherence. To show this, we use a previously reported fluctuation-regulated
quantum master equation [Phys. Rev. A 97, 063837 (2018)].  The fidelity is
found to be a function of the drive-induced dissipative terms as well as the
relaxation terms arising from the qubit-environment coupling; as a result, it
behaves non-monotonically with the drive-amplitude. The existence of an optimum
drive-amplitude implies that the single-qubit operations on open quantum
systems would have an optimal clock speed.

\end{abstract}

\maketitle

\section{Introduction}
Several quantum algorithms have been proved to be computationally superior to their classical
counterparts \cite{dj1992, shor1994algorithms, grover1997, shor1999}. Consequently, the physical
realizations of quantum computers have been one of the major areas of research in the last couple of decades
\cite{gc1997, cirac1995quantum, cory_ensemble_1997, loss_quantum_1998,  ryan_spin_2008, bylander_noise_2011,
yoneda_quantum-dot_2018, huang_fidelity_2019}. The conditions required for the physical realization of
quantum computers have been laid out about two decades ago by DiVincenzo \cite{divincenzo2000physical}. He
argued that the operation time of quantum gates should be much smaller than the timescale of decoherence. 
Moreover, successful implementation of quantum algorithms not only requires the fulfillment of DiVincenzo criteria, but also requires quantum gates of high fidelity to achieve reasonable fault-tolerance. 
As a result, recent years have witnessed significant
improvements in the implementations of high-fidelity gates on various architectures
\cite{yoneda_quantum-dot_2018, bylander_noise_2011, huang_fidelity_2019, ballance_high-fidelity_2016,
song201710}.

Among the recent reports on high-fidelity gates, Ballance \textit{et al.} implemented two-qubit and
single-qubit logic gates using hyperfine trapped-ion qubits driven by Raman laser beams, with fidelity above
99\% for gate times between 3.8 $\mu$s and 520 $\mu$s  \cite{ballance_high-fidelity_2016}. They
experimentally found the maximum gate fidelity for a certain value of the gate time.  To account for this,
the authors attributed the varying fidelity to gate errors. The gate performance has been explained with a
phenomenological error model having a sum of four leading sources of gate errors such as photon scattering
error, motional error, off-resonant error and spin-dephasing error.  In another recent work, Song \textit{et
al.} experimentally generated 10-qubit entangled GHZ state using superconducting circuit with qubit-qubit
interaction mediated by a bus resonator and created a 10-qubit quantum gate with a fidelity of 0.668 $\pm$
0.025 \cite{song201710}.  Also, Huang \textit{et al.} reported two-qubit randomized benchmarking with an average
Clifford gate fidelity of 94.7\% and an average controlled-rotation fidelity of 98\% on silicon-based
quantum dots \cite{huang_fidelity_2019}.  

Bertaina \textit{et al.} experimentally observed the decay of Rabi oscillations of spin
qubits based on rare-earth ions and reported that the decay rate was found to depend on the drive
(microwave) power \cite{bertaina_rare-earth_2007}. Similar drive-induced decoherence (henceforth referred
to as DiD) has earlier been observed experimentally in a variety of systems \cite{devoe_experimental_1983,
boscaino_anomalous_1990, boscaino_non-bloch_1993, nellutla_coherent_2007}. 

Recently, Chakrabarti \textit{et al.} formulated a quantum master equation regulated by
explicitly introduced thermal fluctuations of the environment which is chosen
to be diagonal in the eigen basis \{$|\phi_j\rangle$\} of the static Hamiltonian of the
environment, represented by, $\mathcal{H}_{\rm env.}(t) = \sum_j
f_j(t)|\phi_j\rangle\langle\phi_j|$, where $f_j(t)$-s are assumed to be independent, Gaussian,
$\delta$-correlated stochastic variables with zero mean and standard deviation $\kappa$ \cite{cbQME2018}. This
ensures that the fluctuations would destroy the environment coherences, but do not change the
equilibrium population distribution of the environment. To arrive at the regulator from the
thermal fluctuations, a finite propagator is constructed which is infinitesimal in terms of the
system Hamiltonians, but remains finite in terms of the instances of the fluctuation. Further it
has been assumed that the timescale of the fluctuations of the environment is much faster
compared to the timescale with which the system evolves. The usual Born approximation, \textit{i.e.} the
total density matrix of the system and the environment can be factorized into that of the
system and the environment at the beginning of the coarse-graining interval, is also used \cite{petruccione}.
Following these assumptions, a regular coarse-graining procedure is carried out to obtain the
\emph{fluctuation-regulated} quantum master equation (frQME). We note that the approach to
arrive at the frQME is a hybrid one; the system is treated as Markovian whereas the memory of
the environment is retained through the presence of the fluctuation and as such the environment
is non-Markovian.  This frQME predicts simpler forms of DiD, which have been shown to be the
absorptive Kramers-Kronig pairs of the well-known Bloch-Siegert and light shift terms. The
predicted nature of DiD from the frQME has been verified experimentally \cite{cbEcho2018}. In
the present study, we aim to show that the fast gate operations and achieving high fidelity may
not be two independent processes.  We show that the competition between the two sources of
decoherence, namely, qubit-environment coupling and second order effects of the drive, naturally
leads to an optimum value of the speed of a single-qubit gate. 

It may also be possible to arrive at the similar terms by using polaron and variational polaron
transformations, but the frQME is chosen for its inherent simplicity and robust nature; for
example, the frQME does not require rotating wave approximation (RWA) since the fluctuations
provide all necessary regulators for non-resonant terms \cite{cbQME2018}.  Moreover, the
validity of the frQME requires the drive-amplitude to be small compared to the inverse of the
timescale of the fluctuations. Hence, the frQME can be applied to the cases of strong drive as
long as the above requirement is adequately met. 

\section{Theoretical Preliminaries}
The frQME is given by the following form:
\begin{widetext}
\begin{equation}
\label{frQME}\frac{d}{dt}{\rho_S}(t) = -\frac{i}{\hbar} \; {\rm Tr}_L [H_{\rm eff} (t),{\rho_S}(t)\otimes \rho_L^{\rm eq}]^{sec}\; 
-\frac{1}{\hbar^2} \int_0^\infty d\tau\; {\rm Tr}_L [H_{\rm eff}(t),[H_{\rm eff} (t-\tau),{\rho_S}(t)\otimes \rho_L^{\rm eq}]]^{sec}\; 
e^{-\frac{|\tau|}{\tau_c}}
\end{equation}
\end{widetext}
where $H_{\rm eff}$ contains the drive and the system-environment coupling Hamiltonians,
$\rho_L^{\rm eq}$ denotes the equilibrium density matrix of the environment and $\tau_c$ is the
correlation time of the aforementioned fluctuation \cite{cbQME2018}. The DiD originates from the
double commutator under the integral in the above equation.

To investigate the effect of DiD on quantum gate operations, we apply the frQME on an ensemble
of single-qubit systems coupled to their respective local environments.  As an idealization of
the qubits, we consider spin-1/2 systems having gyromagnetic ratio $\gamma$, and placed in a
static, homogeneous magnetic field $\mathbf{B_{\circ}} = B_{\circ} \hat{k}$ and a resonant
external circularly polarized drive of the form $\mathbf{B_1} = B_1 [\cos(\omega
t+\phi)\;\hat{i} + \sin(\omega t+\phi)  \;\hat{j}] $ (where $\omega$ is the frequency of the
drive, chosen to be equal to the Larmor frequency of the system ($=-\gamma B_{\circ}$) and at
time $t=0$, $\mathbf{B_1}$ makes an angle $\phi$ with respect to $x$-axis) is applied on the
system. An application of the frQME on this system results in the following equation in the
Liouville space,
\begin{equation}
\label{eqLiouville}\frac{d\hat{\rho}_s}{dt} = [-i \hat{\hat{\mathcal{L}}}^{(1)}_{\rm drive} - \hat{\hat{\mathcal{L}}}^{(2)}_{\rm drive} - \hat{\hat{\mathcal{L}}}^{(2)}_{\rm system-env.}]\hat{\rho_s}  
\end{equation}
where $\hat{\hat{\mathcal{L}}}^{(1)}_{\rm drive}$ is the Liouville superoperator or Liouvillian
for the corresponding $-i[H_{\rm drive}, \rho_s]$ term in the master equation. We assume an
isotropic condition such that the system-environment coupling does not appear in the first order
and no cross term between $H_{\rm drive}$ and $H_{\rm system-env.}$ survives the ensemble
average. $\hat{\hat{\mathcal{L}}}^{(2)}_{\rm drive}$ and $\hat{\hat{\mathcal{L}}}^{(2)}_{\rm
system-env.} $ are the second order drive-induced decoherence and regular relaxation terms,
respectively. With the explicit form of the complete superoperator, $ \Gamma=(-i
\hat{\hat{\mathcal{L}}}^{(1)}_{\rm drive} - \hat{\hat{\mathcal{L}}}^{(2)}_{\rm drive} -
\hat{\hat{\mathcal{L}}}^{(2)}_{\rm system-env.}$), the equation (\ref{eqLiouville}) can be
expressed as follows,
\begin{widetext}
\begin{equation}
\label{eqGamma}\frac{d}{dt} \begin{pmatrix}
\rho_{11} \\ \rho_{12} \\ \rho_{21} \\ \rho_{22}
\end{pmatrix} = \begin{pmatrix}
-\frac{\omega_1^2 \tau_c}{2}-\frac{1-m}{T_1} & \xi & \xi^\star & \frac{\omega_1^2 \tau_c}{2} +
\frac{1+m}{T_1} \\
-\xi^\star & -\frac{\omega_1^2 \tau_c}{2}-\frac{2}{T_2} & \eta^\star & \xi^\star \\
-\xi & \eta & -\frac{\omega_1^2 \tau_c}{2}-\frac{2}{T_2} & \xi \\
\frac{\omega_1^2 \tau_c}{2}+\frac{1-m}{T_1} & -\xi & -\xi^\star & -\frac{\omega_1^2
\tau_c}{2}-\frac{1+m}{T_1}
\end{pmatrix} 
\begin{pmatrix}
\rho_{11} \\ \rho_{12} \\ \rho_{21} \\ \rho_{22}
\end{pmatrix} .
\end{equation}
\end{widetext}

The above dynamical equation contains three types of terms, \textit{viz.} (i) the first order nutation
terms, given by $\xi = ie^{i\phi}\omega_1/2$, where $\omega_1 = -\gamma B_1$ is the drive-amplitude in the
angular frequency units; the axis of the drive may be chosen by suitably adjusting $\phi$, (ii) the second
order DiD terms in the diagonal and in the anti-diagonal, given by $\omega_1^2\tau_c$ and $\eta = e^{2i\phi}
\omega_1^2 \tau_c/2$, (iii) the second order relaxation terms which include $T_1$ and $T_2$ to denote
longitudinal and transverse relaxation times, respectively and $m$ which is equal to ${\rm
Tr}(\sigma_z\rho^{\rm eq})$ \cite{cbQME2018}.

To validate the above, we use the equation (\ref{eqGamma}) to analyze a 3-pulse block,
R$_3=\{\pi, -2\pi, \pi\}$, previously described in Chakrabarti \textit{et al.}'s  work
\cite{cbEcho2018}. A straightforward analysis assuming $\omega_1 \gg \frac{1}{T_1},
\frac{1}{T_2}$, shows that after the application of the pulse block, the magnetization is
reduced by a factor of $e^{-(R_{\rm eff}+\omega_1^2 \tau_c)T}$ where $T$ is the duration of the
3-pulse block and $R_{\rm eff}$ denotes $\frac{1}{T_1}+\frac{1}{T_2}$.  We shall use the above
assumption in the subsequent calculations as well, since this does not impose an upper bound on
the drive-amplitude.  In order to arrive at this factor, the equation (\ref{eqGamma}) has been
consecutively solved for three successive pulses and finally the $z$-component of magnetization
has been computed. This expression exactly describes the behavior of the nutating magnetization
of the aforementioned 3-pulse block \cite{cbEcho2018}.

\section{Application of Hadamard Gate on an ensemble of spin-1/2 systems}
We consider the situation of a single Hadamard gate applied on a mixed state written in a pseudopure
form, 
$\rho_0 = \frac{(1-m)}{2} \mathbb{I} +m |0\rangle \langle0|$.
A Hadamard gate is realized by the following unitary propagator, $U_{\rm Hadamard} = e^{i
\frac{\sigma_y}{2}  \frac{\pi}{2}} e^{-i  \frac{\sigma_x}{2} \pi} $ \cite{dmk2003}. The practical
realization of this gate on a single spin qubit will require two square pulses to be applied about $x$-axis and $y$-axis for
duration $\pi/\omega_1$ and $\pi/2\omega_1$, respectively. If the system evolves without any dissipation
during the application of the gate, the final density matrix would be, 
$\rho = \frac{1}{2} (\mathbb{I} + m \; \sigma_x)$.

\begin{figure}[t]
\centerline{\includegraphics[width=0.9\linewidth]{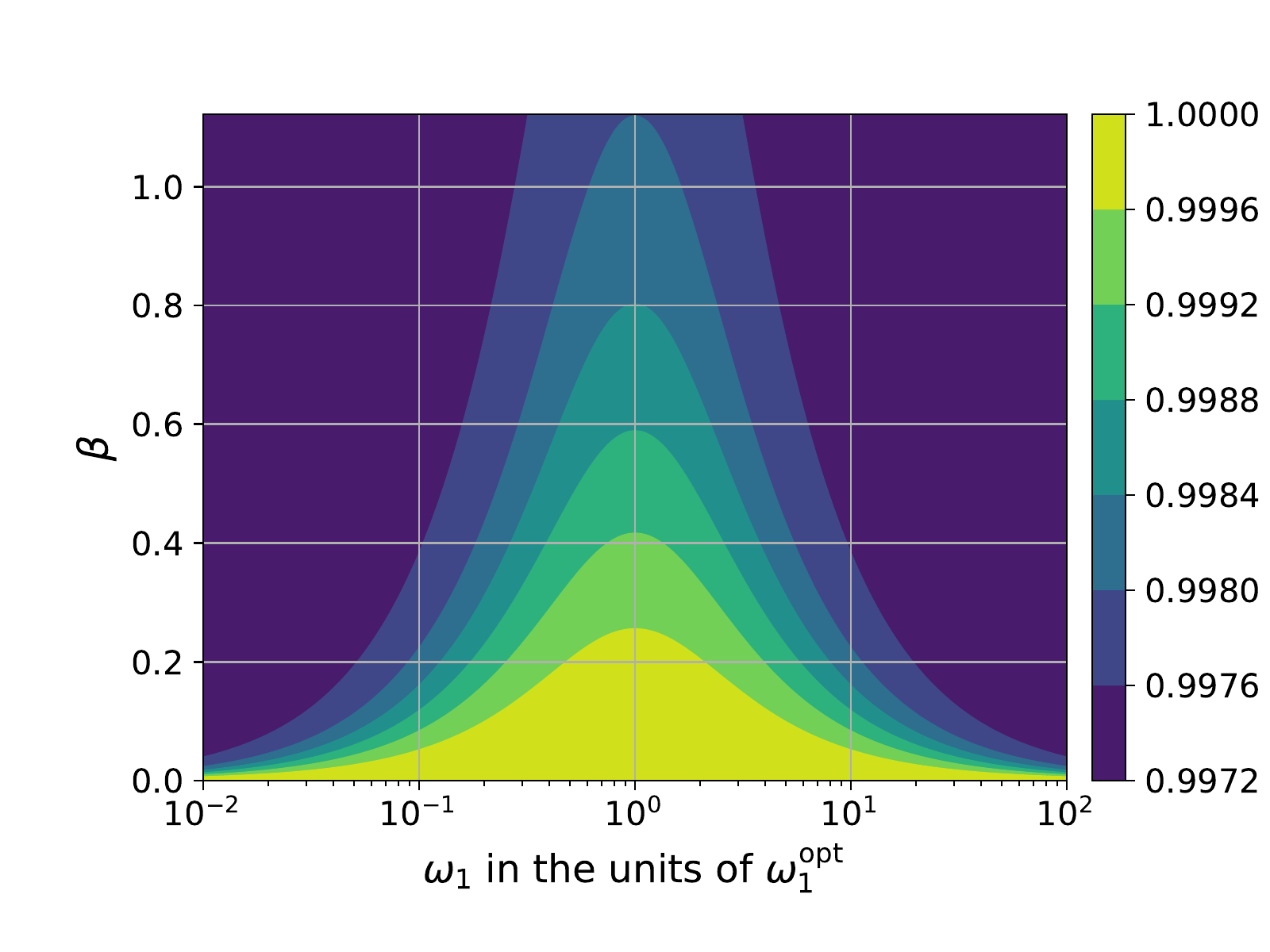}}

\caption{The filled contours show the fidelity as a function of $\omega_1$ and $\beta$. We note that the maximum value of the
fidelity does not depend on $m$; for $\beta\rightarrow 0$, we obtain maximum fidelity to be 1 for all values
of $m$. In this plot, $m$ has been chosen as 0.1.  The central vertical grid at  
$\omega_1 = \omega_1^{\rm opt}$ shows the position of the optimum
drive-amplitude. For a particular value of $\omega_1$, the value of fidelity increases with the
lowering of $\tau_c$, \textit{i.e.} with the lowering of $\beta$.  
We note that $\tau_c \rightarrow 0$ and hence $\beta
\rightarrow 0$, corresponds to the complete absence of all second order terms, in which case, the evolution
of the qubit is unitary.  At this condition, the fidelity is 1 for all values of $\omega_1$  as shown by the
yellow (color online; light gray in print) region at the bottom of the plot. The range of values for which a single color has been used in the plot, are shown in the colorbar on the right with the upper and lower bound of the range.}

\label{fig1}
\end{figure} 

On the other hand, if we consider the dissipative evolution of the system (from the initial $\rho_0$) as dictated by the equation
(\ref{eqGamma}), we obtain a mixed state density matrix, $\rho' = \frac{1}{2} (\mathbb{I} + ma \;
\sigma_x)$ at the end of the gate operation, where 
$a = e^{-\frac{3\pi}{2} \left(\frac{R_{\rm eff}}{\omega_1}
+ \omega_1 \tau_c\right)}$.
To estimate the departure from the unitary
behavior, we calculate the fidelity using the following definition, 
$F(\rho, \rho') = \left({\rm Tr}\sqrt{\sqrt{\rho} \rho' \sqrt{\rho}}\right)^2 $
\cite{uhlmann1976,jozsa1994}.
The fidelity between the expected density matrix $\rho$
and the obtained density matrix $\rho'$, for this particular case, turns out to be,
\begin{equation} \label{eq5}
F = \frac{1}{2} [(1+m^2 a)+\sqrt{(1-m^2)(1-m^2a^2)}].
\end{equation} 
The above form of
fidelity has a maximum value for an optimum value of the drive-amplitude $\omega_1^{\rm opt}$, given by,
\begin{equation}
\label{eq6}\omega_1^{\rm opt} = \sqrt{\frac{R_{\rm eff}}{\tau_c}} .
\end{equation}
  
To show the dependence of the fidelity on the drive-amplitude $\omega_1$ in the unit of $\omega_1^{\rm
opt}$, we rewrite the expression for $a$
as, $a= \exp\left\{ -\beta \left(\omega_1/\omega_1^{\rm opt}~+~\omega_1^{\rm
opt}/\omega_1\right)\right\}$ where $\beta = \frac{3\pi}{2} \sqrt{\tau_c R_{\rm eff}}$.
In figure \ref{fig1}, the filled contours depict the fidelity of the Hadamard gate from the equation
(\ref{eq5}), for
various combinations of $\beta$ and $\omega_1/\omega_1^{\rm opt}$.
The contours show that for smaller $\tau_c$ \textit{i.e.} for smaller $\beta$, the 
fidelity attains a higher value with a wider range of the drive-amplitude. 
In a hypothetical scenario of extreme
motional narrowing, \textit{i.e.} when $\frac{1}{T_1}, \frac{1}{T_2} \sim \omega_{SL}^2 \tau_c $, $\omega_1^{\rm
opt}$ will be of the order of $\omega_{SL}$, where $\omega_{SL}$ is the qubit-environment coupling strength.
Under this condition, the maximum fidelity can be achieved when $\omega_1$ is of the order of $\omega_{SL}$.

To implement the gate operation, one may also use a shaped pulse for which the amplitude, frequency and the phase may be a function of time. Usually these variations are much slower compared to the timescale of the fluctuations. We outline here two approaches to deal with the shaped pulses. In the first approach, a shaped pulse can be constructed as a sequence of narrow square pulses of fixed parameters.
For each of these pulses, one can construct a suitable superoperator $\Gamma$. Sequential application of respective $\Gamma$-s will lead to the final density matrix. This approach is particularly suitable for numerical evaluation of the propagator. In the second approach, which is more suitable for analytical evaluation of the propagator, $\Gamma$ in the equation (\ref{eqLiouville}) can simply be expressed as a function of time since $H_{\rm drive}$ has time-dependent parameters. In such cases, the finite time propagator is given by, $\mathcal{T} \exp\left(\int_{t_{\rm i}}^{t_{\rm f}} dt' \; \Gamma(t')\right) $ where $t_{\rm i}$ and $t_{\rm f}$  indicate the time instances of the beginning and the end of the shaped pulse, respectively and $\mathcal{T}$ is the Dyson time-ordering operator. 


\section{Discussions}
The speed of a classical computer is specified in terms of its clock speed. Similarly, for a
quantum computer, the period of Rabi oscillation or nutation defines the minimum time required for a
single-qubit operation. As such, the frequency of Rabi oscillation ($\omega_1$ in our work) is effectively
the clock speed. So, our result indicates that maximum fidelity can be achieved only for a specific clock
speed which is referred to as the optimal clock speed of the single-qubit gates.  Although we have shown the
existence of the maximum fidelity for the Hadamard gate, the optimality argument can easily be extended to
the other single-qubit gates. All single-qubit gates (except phase gates) involve pulses about the
transverse axis and therefore results in nutation of the targeted qubit. Since the equation (\ref{eqGamma}) is
applicable for generic nutation of the qubits, hence DiD terms are expected to give rise to similar
optimality condition. However, the precise form of the optimum drive-amplitude would vary depending on the
offset frequency of the applied pulse used in the gate operation.

While this work specifically uses the frQME, 
other known forms of DiD such as the fourth order photon-phonon dissipative
terms (second order in drive and second order in qubit-environment coupling) as given by M$\ddot{\rm u}$ller and Stace 
using Keldysh formalism can also be incorporated in this
analysis \cite{keldysh1965, ms2017}. This fourth order DiD term has a form $\omega_1^2
\tilde{T}$ where $\tilde{T}$ is a function of the qubit-environment coupling and other relevant frequencies.
As such, all conclusions drawn above remain valid, although the form of the optimum value of $\omega_1$
would change.

It may appear counter-intuitive that a fluctuation acting on the environment helps regulate even
the drive-drive dissipator. We note that the origin of the regulation lies in the construction
of the finite propagator. An infinitesimal propagator in drive has finite propagator in terms of
the fluctuations. Such a propagator captures the evolution of the environment (under
fluctuation) while the system evolves under the drive. The environment acquires a memory kernel
due to finite time of evolution. Therefore, all second order terms in the construction of the frQME
contain the memory kernel of the environment irrespective of the Hamiltonian appearing in the
second order. We note that this effect appears only if we perform a partial measurement (only on
the targeted system and not the environment) and is mathematically captured by a partial trace over the environmental degrees of freedom.
Further details of this propagator and its implications are discussed in detail in the work of
Chakrabarti and Bhattacharyya \cite{cbQME2018}.

Earlier, Plenio \textit{et al.} obtained an optimal range of the drive power of laser-driven trapped-ion
systems using the following arguments \cite{plenio_realistic_1996}. On such systems, lower drive power
results in longer computation time during which the system may undergo a spontaneous emission. On the other
hand, too high laser power leads to ionization. As such, one obtains a small window of laser power for
practical operating considerations of quantum information processing on such systems. Later, they showed
that even for laser power which was not strong enough to cause ionization, there might be leakages to other
atomic levels because of the laser field (off-resonant) acting between the original and the leak-level
\cite{plenio_decoherence_1997}. They found that such leakages did not have any dependence on the laser
power.  We note that their analysis is confined to multi-level trapped-ion systems and any mechanism similar
to DiD is not taken into account.  On the contrary, we consider two-level systems without any leakage (no
additional levels) and show that DiD and qubit-environment coupling can lead to an optimality condition.

Recently, very high fidelity of quantum gates in silicon-based quantum dot systems have been
reported \cite{yoneda_quantum-dot_2018,huang_fidelity_2019}. Such reports are confined to single
or two silicon-based quantum dots and not an ensemble of quantum systems surrounded by local
environment. Therefore, the notion of local environment at equilibrium, in such systems, may not
be strictly valid. So, our treatment needs to be suitably modified to apply on such systems.  On
the other hand, in the usual settings of ensemble quantum computation, \textit{e.g.} NMR
spectroscopy, the usual values of the drive-amplitude is much smaller than the strength of the
qubit-environment coupling.  For example, the value of $\tau_c$ in solution-state NMR
spectroscopy is typically in the range of pico- to nano-seconds. For a choice of drive-amplitude
of 10 kilo-radian/s and $\tau_c$ of 10 ps, $\omega_1^2\tau_c$ evaluates to $10^{-3}$ Hz and
hence is quite small compared to $1/T_1, 1/T_2 \sim 1$Hz \cite{cbEcho2018}.  But, at low
temperature and in other physical methods, such terms are not negligible and have been
experimentally observed \cite{bertaina_rare-earth_2007, devoe_experimental_1983,
boscaino_anomalous_1990, boscaino_non-bloch_1993, nellutla_coherent_2007}.

It is evident that the clock speed of the highest-fidelity single-qubit operations has an optimum value. For
multi-qubit systems, it is known that the time required for an arbitrary transitional-selective pulse and
hence the overall operation time of a specific task in quantum computation on a multi-qubit network, is
limited by the strength of the qubit-qubit coupling ($J$) \cite{steffen_toward_2001}.  To be adequately
selective, a square pulse must have a duration ($\tau_p$) which is inversely proportional to $J$,
\textit{i.e.} $\tau_p \gtrsim \frac{1}{J}$. This in turn indicates that the drive-amplitude $\omega_1$ must
be less than or, of the order of $J$ (keeping the flip angle constant). Therefore, to achieve maximum
fidelity on such a multi-qubit system, one must satisfy the following condition, $\omega_1^{\rm opt} \sim
\omega_{SL} \leqslant J$. Such a restriction may not be achievable for an architecture based on nuclear
spins, but may be engineered in the quantum dots or superconducting flux qubits.

\section{Conclusions}
We have shown that the speeding up of gate operations on a single qubit by increasing the
drive-amplitude may have detrimental effects on the fidelity of the desired operation. There are two
competing processes which affect the fidelity, \textit{viz.} relaxation from the qubit-environment coupling
and the DiD. For drive-amplitude much lower than the optimum value, the relaxation
terms dominate and the increase in the amplitude of the drive ($\omega_1$) results in faster gate operation
with higher fidelity. In contrast, for drive-amplitude greater than the optimum value, the DiD
processes dominate and reduce the fidelity. Therefore, an optimum value for $\omega_1$ exists for which the
fidelity of a quantum operation reaches its maximum. The optimum value of the drive-amplitude is
proportional to the strength of the qubit-environment coupling.  Consequently, faster gate operations with
maximum fidelity would be aided by better isolation of the qubit-network from the environmental influences,
as one expects intuitively.  Finally, we conclude that the competition between the speed of a quantum gate
and its fidelity is an intrinsic feature for open quantum systems.  We envisage that the notion of the
drive-induced decoherence would play important role in realistic implementation of fast and high-fidelity
quantum gates.

\section{Acknowledgments}
The authors wish to acknowledge Martin B. Plenio, Anil Kumar, Siddhartha Lal, and N. S. Vidhyadhiraja for helpful comments and
discussions, and IISER Kolkata for providing the necessary funding.

\end{document}